# Parametric Heating in a 2D Periodically Driven Bosonic System: Beyond the Weakly Interacting Regime

T. Boulier,[1,2,*] J. Maslek,[1,†] M. Bukov,[3,†] C. Bracamontes,[1] E. Magnan,[1,2] S. Lellouch,[4]
E. Demler,[5] N. Goldman,[6] and J. V. Porto[1,‡]

[1]*Joint Quantum Institute, National Institute of Standards and Technology and the University of Maryland,
College Park, Maryland 20742, USA*
[2]*Laboratoire Charles Fabry, Institut d'Optique Graduate School,
CNRS, Université Paris-Saclay, 91127 Palaiseau cedex, France*
[3]*Department of Physics, University of California, Berkeley, California 94720, USA*
[4]*Laboratoire de Physique des Lasers, Atomes et Molcules, Université Lille 1 Sciences et Technologies,
CNRS; F-59655 Villeneuve d'Ascq Cedex, France*
[5]*Department of Physics, Harvard University, Cambridge, Massachusetts 02138, USA*
[6]*Center for Nonlinear Phenomena and Complex Systems, Université Libre de Bruxelles,
CP 231, Campus Plaine, B-1050 Brussels, Belgium*

We experimentally investigate the effects of parametric instabilities on the short-time heating process of periodically driven bosons in 2D optical lattices with a continuous transverse (tube) degree of freedom. We analyze three types of periodic drives: (i) linear along the $x$-lattice direction only, (ii) linear along the lattice diagonal, and (iii) circular in the lattice plane. In all cases, we demonstrate that the Bose-Einstein condensate (BEC) decay is dominated by the emergence of unstable Bogoliubov modes, rather than scattering in higher Floquet bands, in agreement with recent theoretical predictions. The observed BEC depletion rates are much higher when shaking along both the $x$ and $y$ directions, as opposed to only $x$ or only $y$. We also report an explosion of the decay rates at large drive amplitudes and suggest a phenomenological description beyond the Bogoliubov theory. In this strongly coupled regime, circular drives heat faster than diagonal drives, which illustrates the nontrivial dependence of the heating on the choice of drive.

## I. INTRODUCTION

An area of increasing interest in ultracold atoms concerns the engineering of novel states of matter using highly controllable optical lattices [1]. In this context, a promising approach relies on applying time-periodic modulation to the system, in view of designing an effective time-independent Hamiltonian featuring the desired properties [2–4]. This *Floquet engineering* has emerged as a promising and conceptually straightforward way to expand the quantum simulation toolbox, enabling appealing features such as suppressed [5,6] or laser-assisted [7] tunneling in optical lattices, enhanced magnetic correlations [8], state-dependent lattices [9], and subwavelength optical lattices [10], as well as synthetic dimensions [11,12], synthetic gauge fields [13,14], and topological band structures [15].

Despite these promising applications, progress in Floquet engineering for interacting systems has been hindered by heating due to uncontrolled energy absorption. This heating, which stems from a rich interplay between the periodic drive and interparticle interactions, is a particularly challenging problem in interacting systems, where it is known to occur due to the proliferation of resonances between many-body Floquet states, not captured by the inverse-frequency expansion [4,16]. This problem constrains the applicability of Floquet engineering to regimes where heating is slower than the engineered dynamics [9,17–21]. A deeper understanding of the underlying processes is essential to determine stable regions of the (large) parameter space, where the system is amenable to Floquet engineering. Additionally, interaction-mediated heating is itself an interesting nontrivial quantum many-body process. Energy absorption and entanglement production in periodically driven systems have recently been the focus of theoretical studies [16,22–29] and experimental investigations [5,9,20,21]. It was predicted that, whenever the drive frequency is larger than all single-particle energy scales of the problem, heating succumbs to a stable long-lived prethermal steady state, before it can occur at exponentially long times [25,30–35]. However, it is unclear whether this physics is accessible in interacting bosonic experiments.

A perturbative approach to understanding drive-induced heating is to analyze the underlying two-body scattering processes using Fermi's golden rule (FGR) [21,23,36–38]. In the weakly interacting limit, interactions provide a small coupling between noninteracting Floquet states. However, Floquet states cannot be treated as noninteracting when the Floquet-modified *excitation spectrum* is itself unstable [22,24–26]. These instabilities indicate that heating can occur on a shorter timescale than expected from the scattering theory alone.

For Bose-Einstein condensates (BECs) in optical lattices, increased decay rates arise due to the emergence of unstable collective modes. The resulting parametric heating can be described using a Floquet–Bogoliubov–de Gennes (FBdG) approach [24], and the short-time dynamics is dominated by an exponential growth of the unstable excited modes in the BEC. The depletion time of the condensate fraction provides an experimental window to observe this and related effects. Qualitatively different behavior is expected between scattering and parametric instability rates, most notably, different power laws as a function of the interaction strength, tunneling rate, and drive amplitude.

We experimentally explore these predictions in a 2D lattice subject to 1D and 2D periodic drives, by measuring the decay of the BEC condensed fraction. We provide strong experimental evidence that parametric instabilities dominate the short-time dynamics over FGR-type scattering processes, which are responsible for long-time thermalization [21]. Our experiment reveals effects beyond Floquet-Bogoliubov predictions and points out limitations in the applicability of the FBdG theory.

The experiments are performed on a BEC of $^{87}$Rb atoms loaded into a square 2D optical lattice [39,40] with principal axes along $x$ and $y$, formed by two pairs of counterpropagating laser beams with a wavelength of $\lambda = 814$ nm. The total atom number is $N \simeq 10^5$ ($\pm 20\%$ systematic uncertainty). Two piezoactuated mirrors [41] sinusoidally translate the lattice along $x$ and $y$ with any desired amplitude, relative phase, and angular frequency: $\mathbf{r}(t) = \{\Delta x \sin(\omega t), \Delta y \sin(\omega t + \phi)\}$. We consider the effect of three drive trajectories on the decay rate: translation along $x$ only ($\Delta y = 0$), diagonal translation along $x$ and $y$ ($\Delta y = \Delta x$ and $\phi = 0$), and circular translation ($\Delta y = \Delta x$ and $\phi = \pi/2$). Therefore, the driving is 1D ($x$ only) or 2D (diagonal or circular), in a 2D system (2D array of tubes), as shown in Fig. 1. We express the amplitude $\Delta x$ in terms of the drive-induced maximum effective energy offset between neighboring lattice sites in the comoving frame, $K_0 = \Delta E/\hbar\omega$, where $\Delta E = m\omega^2 a \Delta x$, $a$ is the lattice spacing, and $m$ is the $^{87}$Rb mass. The physical displacement is $\Delta x = \hbar K_0/a\omega m$. The lattice depth $V_0$ is held constant during shaking and is measured in units of lattice recoil energy $E_R = h^2/m\lambda^2$. The lattice tunneling energy $\hbar J$ and the interaction strength $g$ are controlled via $V_0$. The value of $J(V_0)$ and $g(V_0)$ are calculated from the band structure, peak atom density, and scattering length (see Appendix B). This calculation results in the following periodically driven Bose-Hubbard Hamiltonian:

$$\hat{H}(t) = \int_z \sum_{i,j} \{-\hbar J[\hat{a}_{i+1,j}^\dagger(z)\hat{a}_{i,j}(z) + \hat{a}_{i,j+1}^\dagger(z)\hat{a}_{i,j}(z) + \text{H.c.}] - \hat{a}_{i,j}^\dagger(z)\frac{\hbar^2 \partial_z^2}{2m}\hat{a}_{i,j}(z) + \frac{\hbar U}{2}\hat{a}_{i,j}^\dagger(z)\hat{a}_{i,j}^\dagger(z)\hat{a}_{i,j}(z)\hat{a}_{i,j}(z)$$
$$+ \hbar\omega K_0[i\sin(\omega t) + \kappa j\sin(\omega t + \phi)]\hat{a}_{i,j}^\dagger(z)\hat{a}_{i,j}(z)\}, \quad (1)$$

where $\hat{a}_{ij}^{(\dagger)}(z)$ is the annihilation (creation) operator at lattice site $(i,j)$ and transverse position $z$ and $\kappa = 0$ for $x$-only and $\kappa = 1$ for 2D drives. The interaction $U$ is defined such as $U/V\sum_{i,j}\int_z \langle \hat{a}_{i,j}^\dagger(z)\hat{a}_{i,j}(z)\rangle = g$ with $V$ the volume of the system.

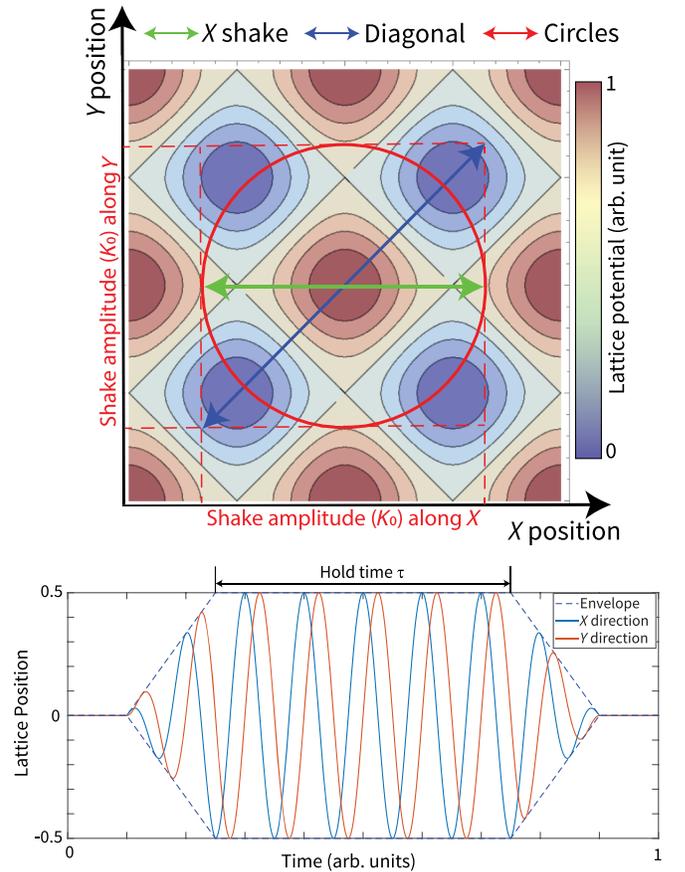

FIG. 1. Schematic of the lattice driving. (Top) Lattice translation is performed along the $x$ direction (green), along the diagonal (blue), and in circles (red). The normalized drive amplitude projected along the $x$ axis, $K_0$, is used to characterize the drive strength for all trajectories. (Bottom) Example of a periodic drive for circular driving.

In order to avoid micromotion effects during a drive period [10], the experiments are performed at integer multiples of the period $T = 2\pi/\omega$ (see Appendix A). The drive amplitude is ramped up smoothly [10,42,43] in a fixed time (minimum 2 ms) corresponding to an integer number of periods (Fig. 1). The shaking is then held at a constant amplitude for a time $\tau$. Finally, the amplitude is ramped down to zero in a few periods. We check the effect of adiabaticity of this procedure (or lack thereof) for the three drive trajectories and describe it in the Appendix A. Once the lattice is at rest, we turn it off in 300 $\mu$s to determine the atomic distribution. We use absorption imaging after the time of flight to measure the condensate fraction as a function of $\tau$.

For most conditions, the condensate decay agrees with an exponential decay [$N(t) = N(0)e^{-\Gamma_{cf}t}$], whose rate $\Gamma_{cf}$ we extract from a least-square fit (see Appendix C). We measure $\Gamma_{cf}$ for the three drives at different values of $\omega$, $K_0$, and $V_0$. FBdG predicts an undamped parametric instability, characterized by exponential growth of unstable modes, i.e., *accelerated* condensate loss. This behavior is inconsistent with the measured exponential decay of the BEC. Hence, the undamped FBdG regime does not last long compared to the typical BEC lifetime for our parameters, and interactions between the excited unstable modes and the BEC play a significant role in the observed heating process. Nonetheless, as we discuss below, the magnitude and scaling of $\Gamma_{cf}$ are well captured by a FBdG description.

Since the Floquet-renormalized hopping is $J_{\text{eff}} = J\mathcal{J}_0(K_0)$ [$\mathcal{J}_\nu(K_0)$ is the $\nu$th order Bessel function], $J_{\text{eff}} < 0$ for $K_0 > 2.4$ and the lowest Floquet band is inverted [5]; the BEC then becomes dynamically unstable at $\mathbf{q} = (0,0)$ [44], but a stable equilibrium occurs at the band edge, and a sudden change in the stability point can trigger a dynamical transition [45] between the two equilibrium configurations. In the band-inverted regime, the stable quasimomenta are $\mathbf{q} = (\pm\pi, 0)$ for a 1D drive along $x$ and $\mathbf{q} = (\pm\pi, \pm\pi)$ for a 2D drive along $x$ and $y$, where the components of the crystal momentum $\mathbf{q}$ are measured in units of the inverse lattice spacing $a^{-1}$. FBdG assumes an initial macroscopic occupation of these modes [24]. Unless stated otherwise, for data taken at $K_0 > 2.4$, we first accelerate the BEC to the appropriate stable point while simultaneously turning on the Floquet drive (see Appendix B).

## II. RESULTS

### A. Lattice depth scans: $\Gamma_{cf}(V_0)$

A major difference between FGR and the FBdG theory is the scaling of the decay rate with the hopping $J$ and interaction strength $g$. Whereas FGR predicts $\Gamma_{cf} \propto (gJ)^2$, the parametric instability rate is expected to be linear ($\Gamma_{cf} \propto gJ$) [24]. Figure 2 shows the condensed fraction decay rate $\Gamma_{cf}$ measured at different $V_0$ (between $7.3E_R$ and $17.0E_R$) and $\omega$, plotted as a function of $gJ/\omega$, for the 2D-diagonal drive and 1D $x$-only drive. The solid lines show the FBdG theory, and the dashed lines are linear fits to the data. For the $x$-only drive, the magnitude and slope extracted from the data are well described by the FBdG theory, and while some deviations appear for the 2D drive, both results are clearly inconsistent with a quadratic dependence.

### B. Amplitude scans: $\Gamma_{cf}(K_0)$

Figure 3 shows the decay rate as a function of the drive amplitude $K_0$, at lattice depth $V_0 = 11E_R$ (which gives $J = 2\pi \times 50$ Hz, $g = 2\pi \times 700$ Hz, and a gap $\Delta E_v/h = 21$ kHz to the next vibrational level) and a drive frequency $\omega = 2\pi \times 2.5$ kHz. The same data are shown on a full range (top) and enlarged (bottom).

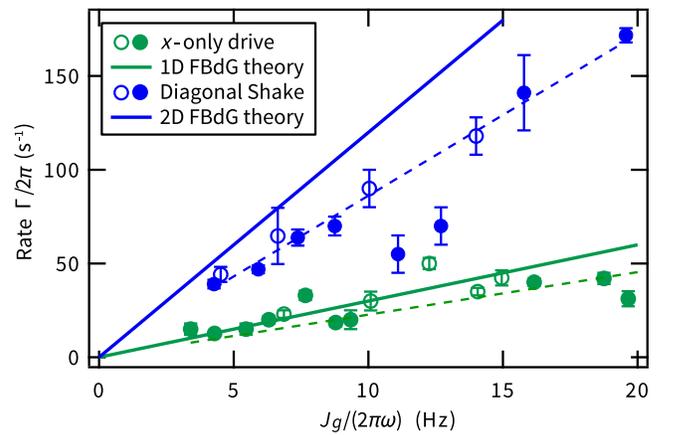

FIG. 2. $\Gamma_{cf}(Jg/\omega)$: Measured decay rates for $K_0 = 2.1$, $\omega = 2\pi \times 4$ kHz (filled circles), and $\omega = 2\pi \times 2.5$ kHz (empty circles) and various lattice depths, compared to the FBdG theory. The horizontal axis is $Jg/(2 \times \omega)$, since the FBdG theory predicts the mode growth rate to be linear in $J$, $g$, and $1/\omega$, as derived in the Appendix E. The dashed lines are linear fits to the data.

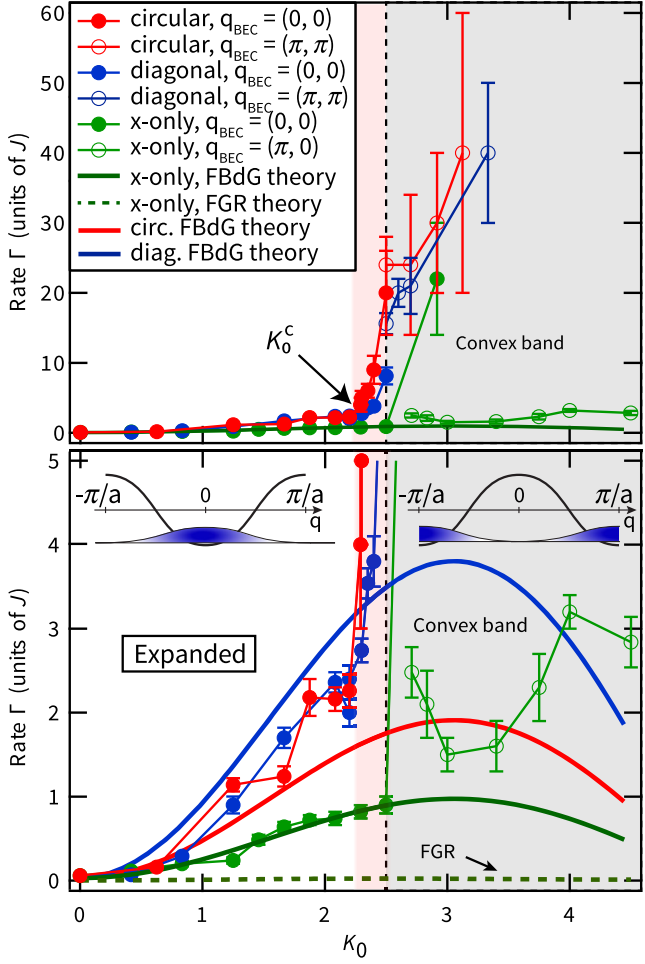

FIG. 3. $\Gamma_{cf}(K_0)$: Comparison between circular (red), diagonal (blue), and x-only (green) drives for $V_0 = 11E_R$ and $\omega = 2\pi \times 2.5$ kHz. The rates are in units of $J = 2\pi \times 50$ Hz, shown full scale (top) and enlarged (bottom). The Floquet–Bogoliubov–de Gennes theory [24] (see Appendix E) $\Gamma_{mum}$ is shown for each drive as solid lines, and the 1D FGR-based scattering theory is shown as the green dashed line (enlarged only). Filled circles indicate data taken at $\mathbf{q} = (0, 0)$, while open circles indicate data taken at $\mathbf{q} = [\pi, \pi(0)]$ (see the text) to keep the BEC in the stable region of the band (illustrated with the bottom plot cartoon). A dramatic increase in the rate occurs at $K_0^c \gtrsim 2.4$ for the circular and diagonal drives, highlighted by the light red region.

Consider first the x-only drive. The instability growth rate predicted from the FBdG theory [16] $\Gamma_{mum} = 8J\mathcal{J}_2(K_0)g/\omega$, which we detail in the Appendix E, agrees with the measured decay rates (Fig. 3). The appearance of the Bessel function $\mathcal{J}_2(K_0)$ in the expression for the rate can be traced back to the parametric resonance condition, requiring the excitation spectrum to match the drive frequency; cf. the Appendix. In contrast, the FGR scattering approach prediction is too low by a factor of about 30, and the predicted scaling, $\propto |\mathcal{J}_2(K_0)|^2$ to leading order, does not describe the data as well. The agreement with the FBdG theory, despite evidence for effects beyond simple undamped parametric instability, is consistent over a range of parameter space in $K_0$, $\omega$, and $V_0$. As expected, the decay dramatically increases when $J_{eff} < 0$ ($K_0 > 2.4$) for $\mathbf{q} = (0, 0)$, while it is partially stabilized when accelerating the BEC to $\mathbf{q} = (\pi, 0)$. We note that significant heating occurs during the drive turn-on and acceleration phase for the $\mathbf{q} = (\pi, 0)$ data, resulting in partial BEC losses.

For the two 2D drives (circular and diagonal), we observe decay rates that follow roughly the same functional form as the 1D drive but about 3× larger. Additionally, the sudden increase in the decay rate $\Gamma_{cf}$ for the 2D drives consistently occurs at a critical amplitude $K_0^c$ below 2.4. At $11E_R$, $K_0^c \simeq 2.15$ (Fig. 3). Above $K_0^c$, both 2D rates increase dramatically beyond any prediction, and the circular rate increases faster than the diagonal rate. This drive dependence and drastic rate increase for $K_0^c < K_0 < 2.4$ suggest effects beyond the FBdG theory, distinct from the simple parametric instability, and are discussed at the end of this paper. For $K_0 \gtrsim 2.4$, the rates are essentially unmeasurable since $\Gamma_{cf} > \omega$. As with the 1D drive, accelerating the BEC to $\mathbf{q} = (\pi, \pi)$ partially stabilizes the decay, just enough to be measurable.

### C. Frequency scans: $\Gamma_{cf}(\omega)$

The FBdG theory predicts distinct behavior at low and high drive frequencies. In the low-frequency regime, the momenta of the maximally unstable mode $\mathbf{q}_{mum}$ evolve as one increases the drive frequency, until it saturates at the Bogoliubov band edge at $\mathbf{q}_{mum} = \{(\pi, 0)\}$, $\{(\pi, 0), (0, \pi)\}$, and $\{(\pi, \pi)\}$ for the x-only, circular, and diagonal drive, respectively. The periodic drive results in interference which depends on the relative phase of the two components of the drive, leading to different most unstable modes for the three shaking geometries. The saturation frequency $\omega_c = E_{eff}^{Bog}(\mathbf{q}_{mum})$ marks the onset of the high-frequency regime [24]. Note that, unlike in 1D lattices [24], the energy of the maximally unstable mode can be lower than the full effective bandwidth (see Appendices B and E). The rate is predicted to increase quasilinearly for $\omega \leq \omega_c$, while $\Gamma_{mum} \propto \omega^{-1}$ for $\omega \geq \omega_c$ [21,24], resulting in a cusp in the rate at $\omega_c$. A detailed derivation of the expressions for the most unstable modes and the instability rates are given in the Appendix.

Figure 4 shows the experimental values for $\Gamma_{cf}(\omega)$ compared with the FBdG theory. Since there is no observed rate explosion for the x-only drive, we use the cusp in $\Gamma_{cf}^x(\omega)$ to calibrate our experimental value for $g$, which agrees to within 20% with an estimate calculated from the lattice parameters (see Appendix B). Using this value of $g$, the prediction for the diagonal drive $\omega_c^{diag}$ matches the experiment. While the FBdG theory predicts the same $\omega_c$ for circular and x-only drives, the measured cusp for the circular drive lies between the cusps of the two linear drives.

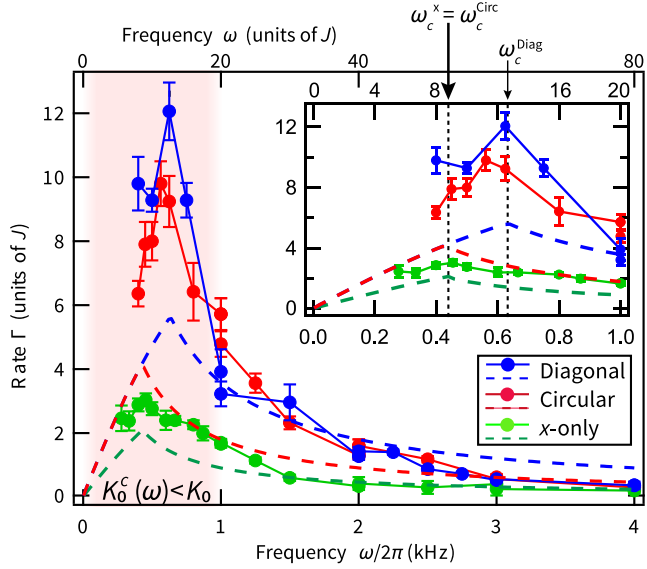

FIG. 4. Decay rates versus drive frequency. $\Gamma_{cf}(\omega)$ for the three drive trajectories at $K_0 = 1.25$ and $V_0 = 11E_R$. The Floquet–Bogoliubov–de Gennes theory $\Gamma_{mum}(\omega)$ is shown for each trajectory as dashed lines. The theoretical cusp positions are marked as vertical black dashed lines. The value we use for $g$ is fitted to the x-only drive cusp and is close to the expected value (see Appendices B and E). A rate explosion occurs at low frequencies when $K_0 > K_0^c(\omega)$ (cf. Fig. 3), represented as the light red zone.

The observed behavior qualitatively fits the FBdG theory, with rates generally higher than predicted for $\omega \sim \omega_c$. For the x-only drive, the measured rates are slightly above the prediction below $2\pi \times 1.5$ kHz, and the agreement is excellent above $2\pi \times 1.5$ kHz, as observed with $\Gamma_{cf}(K_0)$ at $2\pi \times 2.5$ kHz (Fig. 3). The 2D rates show a larger discrepancy at low frequencies and a decent quantitative agreement for $\omega \gtrsim 2\pi \times 2$ kHz. This result is related to the rate explosion appearing for $K_0 > K_0^c$. As we discuss below, the observed value of $K_0^c$ increases with the frequency. This increase implies a similar rate explosion should happen when *decreasing* $\omega$ at fixed $K_0$. This explosion is especially visible with the diagonal drive (Fig. 4): For $\omega < 2\pi \times 1$ kHz, the data abruptly depart from the prediction. This increased rate at low frequencies for 2D drives is likely responsible for the discrepancy between the experiment and theory. The presence of the cusps in the rate explosion region is still expected, since, for $\omega$ low enough, some modes are energetically inaccessible, and the limit $\Gamma_{cf}(\omega \to 0) \to 0$ must be fulfilled.

### D. Rate explosion

Beyond a critical amplitude $K_0^c$, we observe a sudden increase of the 2D-driven decay rate (Fig. 3). The dependence of $K_0^c$ on the frequency for circular driving at $V_0 = 11E_R$ is shown in Fig. 5. We observe that $K_0^c \to 2.4$ as $\omega \to \infty$, suggesting that the giant heating arises from a finite-frequency effect. Because the effect appears for both diagonal and circular drives, whereas a perturbative argument using the inverse-frequency expansion shows that there are finite $1/\omega$ corrections only for the circular drive, we surmise that the effect is likely nonperturbative. Assuming the anomalously strong heating results from an interplay of correlated physics beyond the Bogoliubov regime and recalling that resonance effects beyond the infinite-frequency approximation lead to a $1/\omega$ dependence in the instability rates [2–4], we make the following scaling argument. Viewing the system as an effective Bose-Hubbard model, the strongly correlated regime is reached for $g/J_{eff} \gtrsim 1$. On the other hand, we note that corrections to the infinite-frequency Floquet Hamiltonian scale as $J/\omega$. We make the phenomenological observation that the dimensionless ratio $(g/J_{eff})(J/\omega)$ should be relevant to a combination of beyond-mean-field and finite-frequency effects. When $J_{eff}$ is low in all lattice directions, $(g/J_{eff})(J/\omega)$ is large, and, therefore, these effects should be large. The simple scaling relation $g/\omega \mathcal{J}_0(K_0) = 1$ gives $K_0^c(\omega) = \mathcal{J}_0^{-1}(g/\omega)$ and is shown as the red line in Fig. 5 with the experimental data. The agreement is surprisingly good for such a simple argument. The quantum many-body nature of the rate explosion calls for more extensive study, that promises new insights into periodically driven strongly correlated quantum lattice systems.

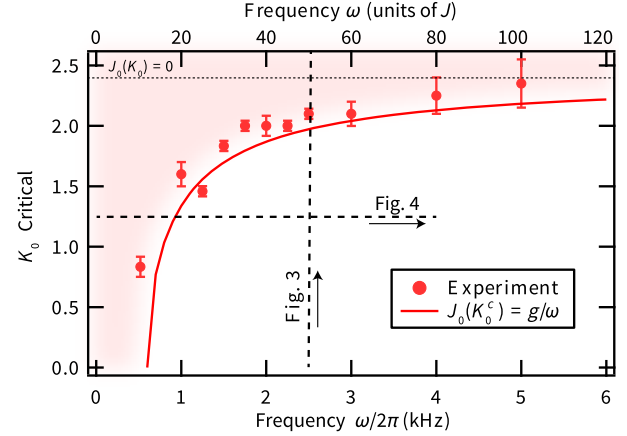

FIG. 5. $K_0^c(\omega)$: Critical drive amplitude measured for various frequencies $\omega$ at $V_0 = 11E_R$. For $K_0 > K_0^c$ (light red region), the decay rate for both 2D drive trajectories increases dramatically, and we observe $K_0^c \to 2.4$ at large $\omega$. A simple relation captures this feature (red line).

We presented a detailed investigation of heating for interacting bosons in a periodically driven 2D lattice. The observed decay rates are substantially larger than expected from a scattering theory based on Fermi's golden rule [21] and scale as expected for interaction-driven parametric instabilities [24]. The observed exponential decay of the condensed fraction suggests that interactions between these excited modes and the BEC, not captured by the FBdG theory, play an important role in the dynamics. Nonetheless, the linear scaling of the condensate loss rate with $gJ/\omega$ is indicative of direct, interaction-induced *instabilities*. Importantly, these instabilities arise from collective modes and involve coherent processes, unlike scattering in a purely FGR approach. In addition, for 2D driving, there exist regions where the heating is even larger than predicted by FBdG, which is not explained by current theories. Altogether, our observations provide important insight into a major heating mechanism in bosonic systems subject to a position-modulation drive, a valuable knowledge for future many-body Floquet engineering schemes.

We note that complementary signatures of parametric instabilities have been recently investigated with bosonic atoms in periodically driven 1D optical lattices [46].

## ACKNOWLEDGMENTS

We thank M. Aidelsburger, I. Bloch, J. Näger, A. Polkovnikov, D. Sels, and K. Wintersperger for valuable discussions. This work was partially supported by the NSF Physics Frontier Center, PFC@JQI (PHY1430094), and the ARO MURI program. Work in Brussels was supported by the FRS-FNRS (Belgium) and the ERC Starting Grant TopoCold. T. B. acknowledges the support of the European Marie Skłodowska-Curie Actions (H2020-MSCA-IF-2015 Grant No. 701034). M. B. acknowledges support from the Emergent Phenomena in Quantum Systems initiative of the Gordon and Betty Moore Foundation, the ERC synergy grant UQUAM, and the U.S. Department of Energy, Office of Science, Office of Advanced Scientific Computing Research, Quantum Algorithm Teams Program. E. D. acknowledges funding from Harvard-MIT CUA (NSF Grant No. DMR-1308435), AFOSR-MURI Quantum Phases of Matter (Grant No. FA9550-14-1-0035), and AFOSR-MURI: Photonic Quantum Matter (Grant No. FA95501610323). E. M. acknowledges the support of the Fulbright Program. We used QuSpin [47,48] to perform the numerical simulations. The authors are pleased to acknowledge that the computational work reported on in this paper was performed on the Shared Computing Cluster which is administered by Boston University's Research Computing Services.

## APPENDIX A: THERMALIZATION OF EXCITED ATOMS

The condensate fraction can decay either by direct Floquet-driven loss or by heating due to relaxation of energetic excitations. The latter mechanism occurs over a thermalization timescale and can be probed by observing relaxation of out-of-equilibrium states in an undriven, static lattice. Special attention to the drive turn-off is required to avoid unwanted excitation due to micromotion during a drive period. Abruptly turning off the drive induces a kick

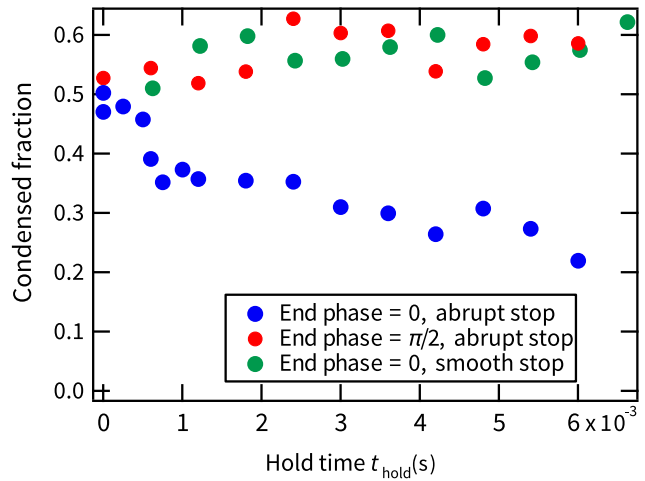

FIG. 6. Thermalization of the nonequilibrium population. The lattice is shaken for a few periods and then held static at $t_{\text{hold}} = 0$. There exists at $t_{\text{hold}} = 0$ a nonequilibrium population that thermalizes with the BEC in a few milliseconds, resulting in a clear decrease of the condensed fraction. The amplitude of this effect depends on the initial size of the nonequilibrium population. It is maximum if atoms are kicked to higher bands by an abrupt stop of the drive (blue). If no such population is created by the stopping of the drive, for example, by choosing an end phase that minimizes the kick (red) or by rapidly ramping down the drive amplitude (green), no condensed fraction decay is observed.

large enough to create a significant out-of-equilibrium population by interband excitation.

To determine the relaxation timescale and test for this additional condensate loss mechanism, we measure the evolution of the condensed fraction when holding the atomic cloud in a static lattice, immediately after an abrupt stop of the drive. While the condensed fraction is initially unchanged, upon letting the static system evolve for a time $t_{\text{hold}}$, we observe a subsequent decrease of the condensed fraction as excited atoms thermalize with the rest of the sample. Figure 6 shows an example of thermalization for a diagonal drive at $\omega = 2\pi \times 2$ kHz. When there is sufficient initial excitation (blue data in Fig. 6), we observe a characteristic thermalization time of the order of 2 ms that does not depend on the initial nonequilibrium population.

We observe that the condensed fraction *increases* slightly with the hold time in the cases where minimal atom excitation occurs (red and green data in Fig. 6). We attribute this increase to rethermalization along the tube axis: Entropy added in the degree of freedom (d.o.f.) associated with the lattice direction can be redistributed along the $z$ (tube) axis, which is visible as a slightly decreased temperature in the $x$-$y$ plane.

The amplitude of the condensed fraction decrease is indicative of the energy of the initial nonequilibrium population. We can measure this amplitude by comparing the condensed fraction at $t_{\text{hold}} = 0$ and $t_{\text{hold}} = 6$ ms, a time sufficient for the thermalization to have occurred. Figure 7 shows the remaining condensate fraction after $t_{hold} = 6$ ms as a function of the phase at which the drive is stopped, for an abrupt stop (blue) and for a smooth but rapid (1 period) turn-off (green). When the stop is abrupt, the fraction of atoms excited depends on the end phase: Abruptly immobilizing the lattice can induce an effective force which depends on the lattice velocity immediately prior to immobilization. Therefore, the excited fraction is maximized when $\dot{q}(t)$ is maximally discontinuous (stop phases of $0[\pi]$ rad) and minimized when smooth (stop phases of $(\pi/2)[\pi]$ rad). In all our decay rate data where we smoothly turn off the drive (in at least one period), no turn-off-induced heating is observed.

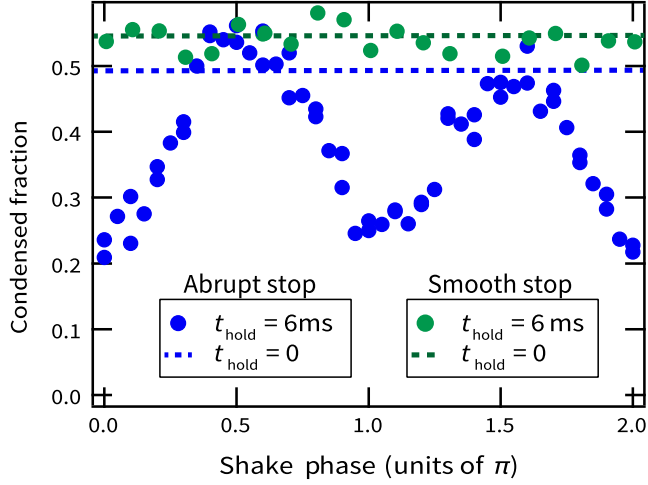

FIG. 7. Amplitude of the thermalization process. The difference between the condensed fraction at $t = t_{end}$ ($t_{hold} = 0$) and at $t = t_{end} + 6$ ms ($t_{hold} = 6$ ms), as a function of the end drive phase, for a $2\pi \times 2$ kHz diagonal (2D) drive. For an abrupt stop (blue), a large population can be transferred out of the BEC for some end phases, which results in a large thermalization event and a clear drop of the condensed fraction. The effect is minimal when stopping the drive such that $q(t)$ is a smooth function. If, however, the drive is stopped by ramping down the amplitude (green), no condensed fraction decay can be observed at any phase.

Abruptly stopping the drive is not the only potential source of nonequilibrium population. The unstable Bogoliubov modes studied in the main text could themselves produce a population that thermalizes and that could potentially modify the measured decay rates. This possibility can be tested with the same stop-and-hold measurement, if turn-off-induced populations are avoided. As visible in Fig. 7, this possibility is realized for stop phases of $(\pi/2)[\pi]$ rad or when smoothly turning off the drive. For our data, avoiding a turn-off-induced population when stopping the drive results in no visible decay, which is shown as the green data in Fig. 6. The result is identical for the whole region of parameter space studied here. We deduce that the energy carried by the nonequilibrium population created by the unstable modes is not enough to significantly impact the measured rates.

## APPENDIX B: EXPERIMENTAL CONSIDERATIONS

*Accelerating the BEC to the band edge.*—In order to accelerate our BEC from $\mathbf{q} = (0, 0)$ to a desired $\mathbf{q} = (q_x, q_y)$, we apply a constant force $\mathbf{F} = \dot{\mathbf{q}}$ for a fixed time in the lattice plane. The force is generated with a constant magnetic gradient, acting on the BEC in the $|F = 1, m_F = -1\rangle$ ground state. Bias coils in the three spatial directions control the gradient direction.

The BEC becomes unstable for quasimomenta about halfway to the band edge, which is the well-known static instability [44]. On the other hand, as mentioned in the main text, ramping up the drive amplitude beyond $K_0 = 2.4$ reverses the band smoothly as $J_{eff}$ becomes negative: $\mathbf{q} \sim (0, 0)$ becomes unstable, and the band edges (or corners, depending on the drive trajectory) become stable. In order keep the BEC in a stable region (in an effectively static dynamical stability sense) at any given time, we synchronize the BEC acceleration with the ramping on of the drive, such that the BEC crosses the static instability point $q = 0.6\pi/a$ when $K_0 = 2.4$.

We accelerate the BEC to $\mathbf{q} = (\pi, \pi)$ for the two 2D drives (diagonal and circle) and to $\mathbf{q} = (\pi, 0)$ for the x-only 1D drive, since these become stable whenever $K_0 > 2.4$.

*Calibration of translation amplitude.*—The piezoactuated mirrors are each roughly calibrated offline using an interferometric technique [41]. The final calibration is performed on the atoms by measuring the tunneling-dependent magnetization decay rate of 2D staggered spin magnetization [49] as a function of drive strength $K_0$. The drive strength at which the tunneling rate $J_{eff}$ vanishes is determined by the condition $K_0 = 2.4$.

*Calibration of tight-binding parameters.*—The lattice depth is calibrated via Raman-Nath diffraction. In the tight-binding limit, the tunneling rate is derived through the modeled 1D dispersion as

$$J \equiv \frac{E(q = \pi/a) - E(q = 0)}{4}. \quad (B1)$$

The on-site interaction $g$ is calibrated from the x-only drive cusp (see Fig. 4). The point at which the rates go from increasing to decreasing, $\omega_c^{diag}$, is given by

$$\omega_c^x = \sqrt{4J_{eff}(4J_{eff} + 2g)}. \quad (B2)$$

Knowing $J$, the experimental value of $\omega_c^x$ offers a calibration for $g$. For $V_0 = 11 E_R$, $J = 50$ Hz and the measured value of $\omega_c^x = 444$ Hz gives $g = 700$ Hz. As an additional check, this value of $g$ is then used to predict the diagonal drive cusp, $\omega_c^{diag} = \sqrt{8J_{eff}(8J_{eff} + 2g)} = 655$ Hz. The observed value of approximately 650 Hz is in good agreement with this prediction.

The interaction strength $g$ depends upon the atom number, the dipole trap, and the lattice parameters. To confirm that the experimental calibration matches these known experimental parameters, we also estimate $g$ through tight binding and Thomas-Fermi assumptions: In the lattice plane, the wave function $\psi(x)$ is taken to be well approximated by a Mathieu function, while we use a Thomas-Fermi profile in the tube direction. The interaction energy $g \propto \iint_{-a/2}^{a/2} |\psi(\mathbf{r})|^4 d^2\mathbf{r}$ is then calculated from the known experimental parameters, including the density profile due to the dipole trap (frequencies $\{\omega_x, \omega_y, \omega_z\} = \{11, 45, 120\}$ Hz). For $V_0 = 11 E_R$, we find $g = 850$ Hz, similar to the calibrated value of 700 Hz. Note that the systematic 20% uncertainty in the atom number can easily explain the small offset between the estimation and calibration.

*Bandwidths.*—It is important to note that $\omega_c$ is, in general, different from the effective bandwidth $B$. For a 2D (diagonal or circular) drive in our 2D lattice,

$$B^{2D} = \sqrt{8J|\mathcal{J}_0(K_0)| \times [8J|\mathcal{J}_0(K_0)| + 2g]} \quad \text{(B3)}$$

and, with a 1D drive in the 2D lattice,

$$B^{1D} = \sqrt{4J[|\mathcal{J}_0(K_0)| + 1] \times \{4J[|\mathcal{J}_0(K_0)| + 1] + 2g\}}, \quad \text{(B4)}$$

which is, in general, different from $\omega_c$, as observed in the main text: Only in the case of the diagonal drive do we find that the maximally unstable mode had the maximum ground band energy, and therefore $\omega_c = B$.

*Background rates.*—All theoretical plots take into account the background decay rate, predominantly due to lattice photon scattering. We experimentally determine this rate by setting $K_0 = 0$ and measuring the resulting rate with the same procedure as in the main text. This constant rate $y_0 \sim 1$ s$^{-1}$ for $V_0 = 11 E_R$ is then added to the FBdG formula for comparison with the experimental data. Since multiphoton resonances to higher bands [50] could complicate the interpretation of the measured decay rates, we perform heating measurements up to drive frequencies of $\omega = 2\pi \times 21$ kHz to identify excited band resonances. Population transfer to higher bands is directly visible on the band-mapping images. The lowest frequency at which resonant higher band excitation is observed is $\omega = 2\pi \times 6.25$ kHz for $V_0 = 11 E_R$, and we therefore limit our heating measurements to below $2\pi \times 4$ kHz to avoid these effects.

*Heating in the absence of a Floquet drive.*—To rule out heating mechanisms that do not depend on the drive but are instead related to the change in $J_{\text{eff}}$, it would be useful to compare the heating observed at a given $K_0$ to the heating observed in a static lattice with a depth chosen to have an equivalent $J = J_{\text{eff}}$ (for $J_{\text{eff}} > 0$). Here, such a direct comparison is experimentally challenging under the same conditions as the Floquet experiment, due to the change in the trap confinement when increasing the lattice depth. In addition to a potential change in gravitational sag (which we minimize by aligning the optical lattice beams directly on the dipole-trapped BEC), the change in trap frequency when changing the lattice depth can excite breathing motion along the nonlattice tube direction, as well as cause a redistribution of atoms within the trap transverse to the lattice direction. These trap-changing effects do not occur for the Floquet modification of the tunneling. To avoid this motion or redistribution, one needs to increase the lattice adiabatically, i.e., on a slower timescale than the Floquet drive is turned on. Nonetheless, we do a type of comparison to the data in Fig. 2 by turning on the lattice slowly (over 200 ms) to a final lattice depth to obtain a given static $J$ and observe the condensate fraction. As an example, we increase the lattice depth to $18.2 E_R$, which changes the tunneling from $J = 2\pi \times 50$ Hz at $11 E_R$ to $J = 2\pi \times 12$ Hz at $18.2 E_R$, corresponding in the Floquet case to $K_0 = 2$. The fact that there is still condensate remaining after the 200 ms turn-on is already an indication that the condensate decay rate in the unshaken lattice is slow. Measuring the condensate decay rate at this depth gives $\Gamma/2\pi = 2$ s$^{-1}$, which is slower than the equivalent Floquet rate by a factor of 9 and slower than all the data at $K_0 = 2.1$ shown in Fig. 2 by at least a factor of 7.

## APPENDIX C: EXTRACTING THE DECAY RATES

*Rate extraction.*—Our data consist of a series of measured condensed fractions after various driving times. A time series typically presents an exponential-looking decay. An example for such a decay is given in Fig. 8. Since we focus on the early time decay rate, greater weight is given to earlier data points. We fit to an exponential with no offset (two fit parameters): $f(t) = Ae^{-\Gamma_{cf}t}$, where $A$ is the $t = 0$ condensed fraction (typically, $A \geq 0.5$). The rates presented in this manuscript are the extracted fit parameter $\Gamma_{cf}$, and the uncertainties correspond to $\pm 1$ standard deviation.

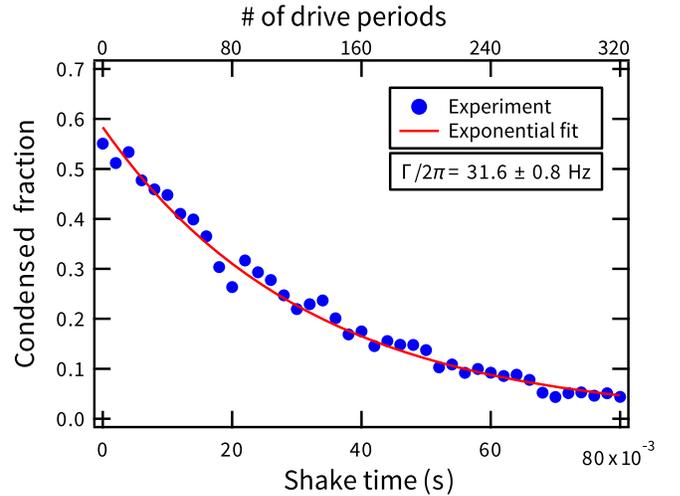

FIG. 8. Typical decay plot. Each data point (blue dots) is an experimental realization, where the condensed fraction is measured after a variable shake time (all else kept constant). The resulting decay curve is then fitted with an exponential function (red thick line). This particular example uses $K_0 = 2$, $\omega/2\pi = 4$ kHz, and $V_0 = 11 E_R$. Each experimental rate in the manuscript is extracted from such a fit, and its error bar is given by $\pm 1$ standard deviation.

A small fraction of the experimental data does not show an exponential decay. In such cases, the condensed fraction versus time looks linear, which is indicative of an *acceleration* relative to an exponential decay. Whenever fitting to an exponential is impossible, we take the slope, focusing on early times where the condensed fraction is more than half that of $t = 0$. In any case, the measured decay rate under all conditions corresponds to the decay rate at initial times.

For these rates to be directly compared to the BdG prediction, additional considerations must be taken. First, let us consider a maximally unstable Bogoliubov mode with an *amplitude* predicted to grow as $e^{\Gamma_{mum}t}$. An experiment will actually detect a rate $2\Gamma_{mum}$, as it measures an amplitude *squared* (typically, the number of atoms in the unstable mode). Second, since the experiment measures how many atoms leave the BEC (to populate the modes) per unit of time, it is sensitive to the number of simultaneous maximally unstable modes, as each is a decay channel. If two modes are equally and maximally unstable, as is possible in 2D, then an additional factor of 2 is needed in the theory to compare to the experiment. This multiple-modes factor is 1 for the $x$-only drive and 2 for the circle and diagonal drives. All these extra factors are added to the theory plots throughout this paper: In total, the $x$-only drive theory is $2\times$ and the circle and diagonal drives are $4\times$ larger than the bare BdG rates predicted in Ref. [24]. These points are expanded in the derivation of the FBdG rate $\Gamma_{mum}$ later in this Appendix.

*Rate explosion: Circular versus diagonal.*—We observe that the circular decay rate increases faster than the diagonal rate for $K_0 \gtrsim K_0^c$. For example, Fig. 9 shows an enlarged version of Fig. 3, where $V_0 = 11E_R$ and $\omega = 2\pi \times 2.5$ kHz, to make this observation clearer. The rates go from being approximately equal below $K_0^c$ to $\Gamma_{cf}^{circ} > 2\Gamma_{cf}^{diag}$ when tunneling is suppressed. Since breaking time-reversal symmetry is necessary for many proposed schemes, this observation may be of interest to the Floquet engineering community.

*Difficulties associated with dynamical rates.*—In the main text, we compare the decay rates measured in the experiment to the instability rates predicted by the FBdG theory. Here, we elaborate on some intrinsic difficulties in the procedure which may affect the extracted values.

As explained in Ref. [24], for drive frequencies below the effective drive-renormalized Floquet-Bogoliubov bandwidth, there exists an entire manifold of resonant modes. While all of them contribute to expectation values of observables at very short times, only the maximally unstable mode $\mathbf{q}_{mum}$ dominates the long-time BdG dynamics, and the rate associated with $\mathbf{q}_{mum}$ sets the parametric instability rate. Thus, at any finite time, the FBdG dynamics is in a crossover between these two regimes, which shrinks exponentially with time. Yet the time width of this crossover also depends on the drive frequency: The higher the frequency, the smaller the instability rate, and the longer it takes for the exponential behavior to become visible.

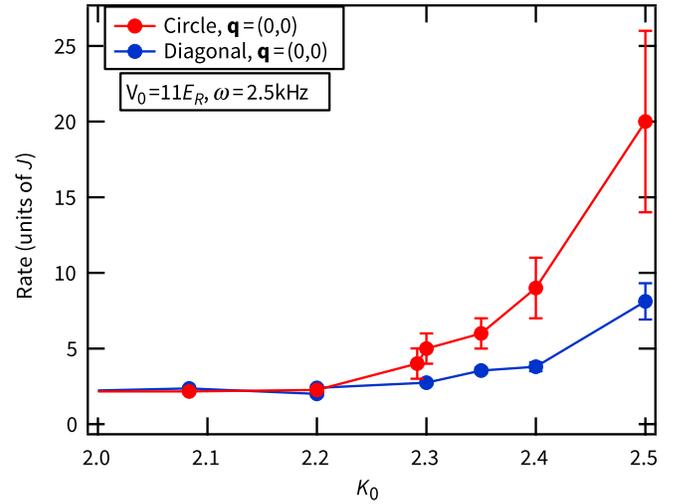

FIG. 9. Decay rates in the critical region. Enlarged plot of the data shown in Fig. 3, focusing on the region between $K_0^c$ and $K_0 = 2.5$. The difference between the circular and the diagonal drive is clearly visible, with the circular drive rate as much as twice that of the diagonal drive rate.

When extracting the rates from data, effects due to this crossover become relevant. To test this, we perform exact numerical simulations of the BdG equations of motion and compute the dynamics of the excited fraction of atoms $n_{\mathbf{q}}(t)$ over a finite number of driving cycles, which increases suitably with the drive frequency. We then extract the instability rates using least-square fitting as the slope of $\log n_{\mathbf{q}}(t)$ *over the last eight driving cycles*, to maximally eliminate transient effects. A comparison between the numerically extracted rates and the analytic theory prediction is shown in Fig. 10 for the three types of drives. Note that the agreement becomes worse at larger $\omega$, since this decreases the rate and pushes the exponential regime to later times. This disagreement is a source of error, which is certainly relevant for the experimental determination of the rates.

Additionally, in the experiment there are strong beyond Bogoliubov effects, not captured by the FBdG theory. Because of the nonlinearity of the Gross-Pitaevskii equation which leads to saturation of the condensate depletion, the exponential BdG regime mentioned above crosses over into a third, scattering-dominated regime. In this regime, the population transferred coherently to the maximally unstable modes by the parametric resonance starts decaying into the surrounding finite-momentum modes at high

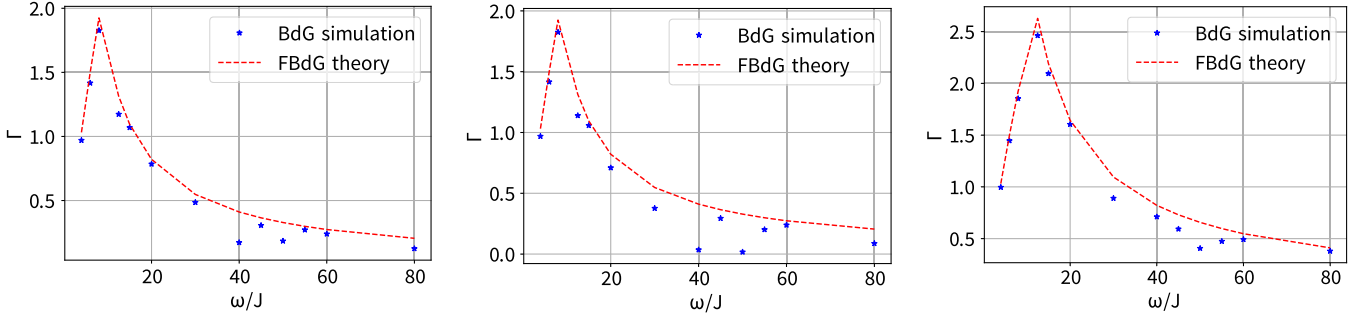

FIG. 10. Rate comparison: FBdG theory versus BdG simulation. Comparison between the dynamically extracted instability rates from solving the exact BdG equations of motion for a finite number of drive cycles (blue stars) and the FBdG prediction (red dashed line). The three plots correspond to *x*-only (left), 2D circular (middle), and 2D diagonal (right) drives. The numerical simulations include a tube transverse d.o.f. The parameters are $g/J = 12$, $K_0 = 1.25$, and a momentum grid of $80 \times 80 \times 101$ modes in the $x$, $y$, and $z$ direction, respectively.

## APPENDIX D: HEATING DYNAMICS OF THE TRUNCATED WIGNER APPROXIMATION

While the FBdG theory is valid in the short-time regime of the dynamics, it has some serious deficiencies. Perhaps the most notable of these, when it comes to out-of-equilibrium dynamics, is the lack of particle-number conservation: The condensate is assumed to be an infinite reservoir which supplies particles to indefinitely increase the occupation of pairs of modes with finite and opposite momenta. In equilibrium, this description works well and captures the physics in the superfluid phase. Away from equilibrium, however, condensate depletion processes such as the parametric instabilities studied in this work lead to the significant depletion of the BEC, and the mean-field Bogoliubov description ultimately breaks down under typical observation times.

Particle conservation is obeyed in the truncated Wigner approximation (TWA), which also includes nonlinear interactions modeling collisions between Bogoliubov quasiparticles [51,52], and is capable of describing thermalization at later stages, due to the continuous pumping of energy into the system.

The starting point for the TWA is the Gross-Pitaevskii equation which, in the comoving real-space frame, reads ($\hbar = 1$)

$$i\partial_t a_\mathbf{r}(t) = -J[a_{\mathbf{r}+a\mathbf{e_x}}(t) - a_{\mathbf{r}-a\mathbf{e_x}}(t)] - J[a_{\mathbf{r}+a\mathbf{e_y}}(t) - a_{\mathbf{r}-a\mathbf{e_y}}(t)] - \frac{\partial_z^2}{2m} a_\mathbf{r}(t)$$
$$+ \omega K_0 \mathbf{r} \cdot [\sin(\omega t)\mathbf{e_x} + \kappa_d \sin(\omega t + \phi)\mathbf{e_y}] a_\mathbf{r}(t) + U|a_\mathbf{r}(t)|^2 a_\mathbf{r}(t), \quad \text{(D1)}$$

where $a_\mathbf{r}(t)$ models the bosonic system at time $t$ and position $\mathbf{r} = (x, y, z)$. The kinetic energy reflects the lattice d.o.f. in the $(x, y)$ plane and the continuous transverse mode along the $z$ axis. The periodic drive is in the $(x, y)$ plane with frequency $\omega$ and amplitude $K_0$. $\kappa_d = 0$ for the *x*-only drive, and $\kappa_d = 1$ for both 2D drives. $\phi$ is the relative drive phase between $x$ and $y$ ($\phi = 0$ for a diagonal drive and $\phi = -(\pi/2)$ for a circular drive). Finally, the on-site interaction strength is denoted by $U$ (such that $U/V \sum_{i,j} \int dz |a_\mathbf{r}|^2 = g$).

Since at time $t = 0$ the system forms a BEC of $N_0$ atoms in a volume $V$, assuming a macroscopic occupation $n_0 = \sqrt{N_0/V}$ in the uniform ($\mathbf{q} = 0$) condensate, the field $a_\mathbf{r}$ can be decomposed as

$$a_\mathbf{r} = n_0 + \frac{1}{\sqrt{V}} \sum_{\mathbf{q} \neq 0} u_\mathbf{q} \gamma_\mathbf{q} e^{-i\mathbf{q} \cdot \mathbf{r}_j} + v_{-\mathbf{q}}^* \gamma_{-\mathbf{q}}^* e^{+i\mathbf{q} \cdot \mathbf{r}_j}, \quad \text{(D2)}$$

where $u_\mathbf{q}$ and $v_\mathbf{q}$ are the Bogoliubov modes which solve the time-independent BdG equations at $t = 0$ [24]. Here, $\gamma_\mathbf{q}$ is a complex-valued Gaussian random variable (associated with the quantum annihilator $\hat{\gamma}_\mathbf{q}$ of Bogoliubov modes) with the mean and variance set by the corresponding quantum expectation values in the Bogoliubov ground state [51].

Hence, in the TWA, one draws multiple random realizations of $\gamma_\mathbf{q}$, each of which corresponds to a different initial state. One then evolves every member of this ensemble according to Eq. (D1), computes the observable of interest, and takes the ensemble average $\overline{(\cdot)}$ in the end. For instance, one can compute the total number of excited atoms as

$$n_{\text{ex}}(t) = \overline{\frac{1}{V} \sum_{\mathbf{q} \neq \mathbf{0}} |a_{\mathbf{q}}(t)|^2}, \quad (D3)$$

which, due to particle number conservation, also reflects the dynamics of the condensate depletion.

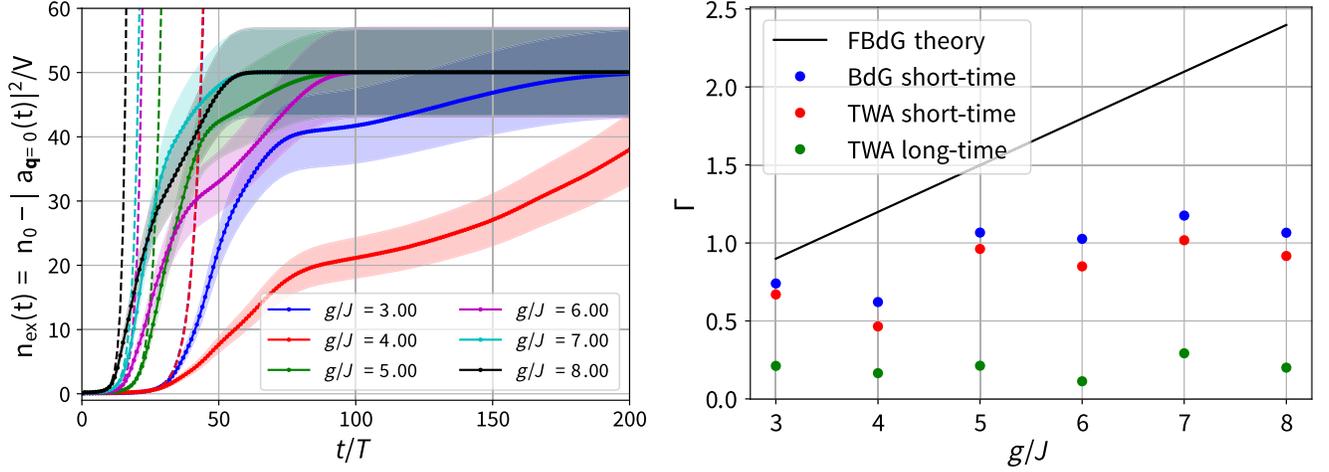

FIG. 11. Rate comparison: TWA versus BdG. Left: Numerical simulation of the excitations growth using TWA (solid lines) and FBdG (dashed lines). The TWA curves change the curvature beyond the regime of validity of BdG. Right: Excitation (i.e., condensate depletion) growth rate against effective interaction strength $g$ for the FBdG theory (black), BdG short-time evolution (blue), TWA short-time evolution (red), and TWA long-time evolution (green). The parameters are $K_0 = 2.1$, $\omega/J = 20.0$, and $n_0 = 50.0$. We use a system of $80 \times 80 \times 101$ momentum modes in the $(x, y, z)$ direction, respectively. The TWA data are averaged over 50 independent realizations, and the error bars (shaded area) are computed using a bootstrapping approach.

In general, we expect that the condensate depletion curve shows two types of behavior: At short times, the FBdG theory applies and $n_{\text{ex}}(t) \sim \exp(2\Gamma_{\text{mum}} t)$ grows exponentially in time. Hence, the condensate depletion curve $|a_{\mathbf{q}=\mathbf{0}}(t)|^2 = V[n_0 - n_{\text{ex}}(t)]$ is concave. At long times, nonlinear interaction effects in the Gross-Pitaevskii equation become important, leading to saturation, and the curvature of condensate depletion changes sign. Therefore, in the long-time regime, the curve is concave. The opposite behavior is true for the evolution of the excitations $n_{\text{ex}}(t)$. The curvature of the experimental data (cf. Fig. 8) suggests that the system enters well into the long-time regime. Yet, the measured decay rates appear consistent with the short-time Bogoliubov theory (main text).

To shed light on this intriguing observation, we perform TWA simulations on a periodically driven homogeneous system in $(2 + 1)$ dimensions and extract the short-time and long-time rates from the numerical data. We use a comparison with the BdG simulations, to separate the short-time regime (where agreement between BdG equations of motion and TWA is expected) from the longer-time regime (Fig. 11, left). For the sake of comparison with experiments, we fit the long-time TWA growth to an exponential, even though we find that it follows a more complicated functional form.

Figure 11 (right) shows a scan of the TWA rates over the effective interaction parameter $g$. We find that both the short-time and long-time rates are of similar strength. More importantly, they do not show a quadratic scaling in $g$, as predicted by Fermi's golden rule. This behavior is consistent with the experimental observations. Note the mismatch between the FBdG theory (black) and the short-time BdG simulations (blue), which arises since the most unstable mode does not yet dominate the dynamics at such short times (see Fig. 10 and the corresponding discussion). Indeed, we find an excellent agreement between BdG numerics and the FBdG theory if we extract the rates from the long-time regime. The short-time BdG rates agree qualitatively with the short-time TWA rates, as expected from the agreement seen in Fig. 11 (left). The rates are extracted from a least-square fit over the last five consecutive driving cycles of the short-time region of agreement between BdG and TWA. Since the rates are dynamical, i.e., change depending on the time window used to extract them, the curves in Fig. 11 (right) are not smooth.

We also do frequency and amplitude scans of the long-time TWA rates to look for signatures of the Bessel function $\mathcal{J}_2(K_0)$ and the cusp at the critical frequency $\omega_c$, as expected from the FBdG theory and found experimentally. Unfortunately, we do not see clear signatures of such behaviors in our TWA simulations. Thus, we cannot conclude that the TWA captures the long-time thermalization dynamics of driven bosonic cold atom systems accurately. More interestingly, the rate explosion (see the main text) is also beyond the TWA dynamics, suggesting that quantum effects, such as loss of coherence, are important for describing this phenomenon. Another possible reason for the disagreement is the single-band approximation, as its validity for the Floquet system has not been fully understood so far.

## APPENDIX E: MOST UNSTABLE MODES FOR LINEAR, DIAGONAL, AND CIRCULAR LATTICE DRIVES WITHIN FBDG THEORY

In this Appendix, we briefly revisit the derivation for the most unstable mode within the FBdG theory [26] and extend the results to diagonal and circular drives. The take-home message is that the critical saturation frequency $\omega_c$, which defines the position of the cusp in the instability rate curves, coincides for the linear and circular drives; for the diagonal drive, the value is twice as large on the square lattice. It is straightforward to extend the analysis below to other lattice geometries.

The starting point is the BdG equations of motion (EOM) for the mode functions $u_{\mathbf{q}}(t)$ and $v_{\mathbf{q}}(t)$ in the rotating frame:

$$i\partial_t \begin{pmatrix} u_{\mathbf{q}} \\ v_{\mathbf{q}} \end{pmatrix} = \begin{pmatrix} \varepsilon(\mathbf{q},t)+g & g \\ -g & -\varepsilon(-\mathbf{q},t)-g \end{pmatrix} \begin{pmatrix} u_{\mathbf{q}} \\ v_{\mathbf{q}} \end{pmatrix}, \quad (E1)$$

where $\varepsilon(\mathbf{q},t) \geq 0$ denotes the time-dependent free lattice dispersion relation (shifted by the chemical potential so that it is non-negative).

To analyze the effects of parametric instabilities, we first isolate the time-average term and write $\varepsilon(\mathbf{q},t) = \varepsilon_{\text{eff}}(\mathbf{q}) + g(\mathbf{q},t)$, which will separate the right-hand side of the BdG EOM into an effective time-averaged term and a time-periodic perturbation. Following Ref. [24], we now apply a phase rotation, followed by a static Bogoliubov transformation with respect to $\varepsilon_{\text{eff}}(\mathbf{q})$:

$$\begin{pmatrix} u_{\mathbf{q}} \\ v_{\mathbf{q}} \end{pmatrix} = \begin{pmatrix} \cosh(\theta_{\mathbf{q}}) & \sinh(\theta_{\mathbf{q}}) \\ \sinh(\theta_{\mathbf{q}}) & \cosh(\theta_{\mathbf{q}}) \end{pmatrix} \begin{pmatrix} e^{-2iE^{\text{Bog}}_{\text{eff}}(q)t} & 0 \\ 0 & 1 \end{pmatrix} \begin{pmatrix} \tilde{u}'_{\mathbf{q}} \\ \tilde{v}'_{\mathbf{q}} \end{pmatrix},$$

where $\cosh(2\theta_{\mathbf{q}}) \equiv [\varepsilon_{\text{eff}}(\mathbf{q}) + g]/E^{\text{Bog}}_{\text{eff}}(\mathbf{q})$ and $\sinh(2\theta_{\mathbf{q}}) \equiv g/E^{\text{Bog}}_{\text{eff}}(\mathbf{q})$ with

$$E^{\text{Bog}}_{\text{eff}}(\mathbf{q}) = \sqrt{\varepsilon_{\text{eff}}(\mathbf{q})[\varepsilon_{\text{eff}}(\mathbf{q}) + 2g]}$$

the time-averaged Bogoliubov dispersion. The Bogoliubov transformation diagonalizes the effective static time-average term, while the phase rotation makes it easier to identify the parametric resonant terms (see Appendix C in Ref. [24] and Supplemental Material in Ref. [25] for an application to the paradigmatic parametric oscillator). The BdG EOM now read

$$i\partial_t \begin{pmatrix} \tilde{u}'_{\mathbf{q}} \\ \tilde{v}'_{\mathbf{q}} \end{pmatrix} = \left[ E^{\text{Bog}}_{\text{eff}}(\mathbf{q})\hat{\mathbf{1}} + \hat{W}_{\mathbf{q}}(t) + \sinh(2\theta_{\mathbf{q}}) \begin{pmatrix} 0 & h_{\mathbf{q}}(t)e^{-2iE^{\text{Bog}}_{\text{eff}}(\mathbf{q})t} \\ -h_{\mathbf{q}}(t)e^{2iE^{\text{Bog}}_{\text{eff}}(\mathbf{q})t} & 0 \end{pmatrix} \right] \begin{pmatrix} \tilde{u}'_{\mathbf{q}} \\ \tilde{v}'_{\mathbf{q}} \end{pmatrix}, \quad (E2)$$

where

$$W_{\mathbf{q}}(t) = \begin{pmatrix} g_{\mathbf{q}}(t)\cosh^2(\theta_{\mathbf{q}}) + g_{-\mathbf{q}}(t)\sinh^2(\theta_{\mathbf{q}}) & 0 \\ 0 & -g_{-\mathbf{q}}(t)\cosh^2(\theta_{\mathbf{q}}) - g_{\mathbf{q}}(t)\sinh^2(\theta_{\mathbf{q}}) \end{pmatrix},$$

$$h_{\mathbf{q}}(t) = \frac{1}{2}[g_{\mathbf{q}}(t) + g_{-\mathbf{q}}(t)]. \quad (E3)$$

By the definition of $g_{\mathbf{q}}(t)$, the diagonal matrix $W_{\mathbf{q}}(t)$ has a vanishing period average and, hence, does not contribute to the parametric resonance to leading order. At the same time, the off-diagonal term in Eq. (E2) will be dominant, whenever $h_{\mathbf{q}}(t)$ interferes with the phase term $e^{2iE^{\text{Bog}}_{\text{eff}}(\mathbf{q})t}$ constructively. This latter condition gives the resonant frequencies. In a sense, $h_{\mathbf{q}}(t)$ can be thought of as an effective periodic drive, the amplitude of which, multiplied by the prefactor $\sinh 2\theta_{\mathbf{q}} = g/E^{\text{Bog}}_{\text{eff}}(\mathbf{q})$, determines properties of the resonance, such as the maximally unstable mode.

Let us now analyze the function $h_{\mathbf{q}}(t)$ for the three types of drives we study in the main text.

*Linear drive.*—Using the Jacobi-Anger identity, we have

$$\varepsilon^{\text{lin}}_{\mathbf{q}}(t) = 4J\left[\sin\frac{q_x}{2}\sin\left(\frac{q_x}{2} - K_0\sin\omega t\right) + \sin^2\frac{q_y}{2}\right],$$

$$h^{\text{lin}}_{\mathbf{q}}(t) = 4J\left[\cos(K_0\sin\omega t)\sin^2\frac{q_x}{2} + \sin^2\frac{q_y}{2}\right] - \varepsilon_{\text{eff}}(\mathbf{q})$$

$$= 8J\sin^2\frac{q_x}{2}\sum_{l=1}^{\infty} \mathcal{J}_{2l}(K_0)\cos(2l\omega t).$$

The fact that the system is driven only along the $x$ direction is reflected in the momentum dependence of $h_{\mathbf{q}}(t)$. Applying the rotating wave approximation to the time-periodic off-diagonal term in Eq. (E2), we conclude that the dominant resonant harmonic is $l = 1$, which gives $\omega = E_{\text{eff}}^{\text{Bog}}(\mathbf{q})$. Using the relation $\sinh 2\theta_{\mathbf{q}} = g/E_{\text{eff}}^{\text{Bog}}(\mathbf{q}) = g/\omega$, the momentum-dependent instability rate reads [24]

$$s_{\text{lin}}(\mathbf{q}) = 4J \mathcal{J}_2(K_0) \sin^2\left(\frac{q_x}{2}\right) \frac{g}{\omega}. \quad (E4)$$

Clearly, there are many resonant modes which satisfy the condition $\omega = E_{\text{eff}}^{\text{Bog}}(\mathbf{q}_{\text{res}})$. This condition is visualized as a plane, parallel to the $(q_x, q_y)$ plane, cutting through the dispersion relation: The resulting set of modes defines the solutions to the resonant condition.

However, out of all resonant modes, not all have the same instability rate, because $s_{\text{lin}}(\mathbf{q})$ is a function of $\mathbf{q}$. We are interested in the long-time BdG dynamics, which is dominated by the most unstable mode:

$$\gamma = \max_{\mathbf{q}} s(\mathbf{q}), \qquad \mathbf{q}_{\text{mum}} = \operatorname{argmax}_{\mathbf{q}} s(\mathbf{q}). \quad (E5)$$

Since $s_{\text{lin}}(\mathbf{q})$ is a monotonic function of only $q_x$, the most unstable modes are those resonant modes which have the largest $q_x$ component. Gradually increasing the drive frequency $\omega$, one reaches a critical saturation frequency $\omega_c = E_{\text{eff}}^{\text{Bog}}(\pi, 0)$ at the edge of the Brillouin zone. Using the same analysis as in Ref. [24] and keeping in mind the existence of a continuous d.o.f. in the $z$ (tubes) direction, one can show that, for $\omega \geq \omega_c^{\text{lin}}$, we have $\mathbf{q}_{\text{mum}}^{\text{lin}} = (\pi, 0)$ and $\gamma_{\text{lin}} = 4J\mathcal{J}_2(K_0)g/\omega$, while for $\omega \leq \omega_c^{\text{lin}}$, we find $\mathbf{q}_{\text{mum}}^{\text{lin}} = 2\arcsin\sqrt{[\sqrt{g^2 + \omega^2} - g/4J\mathcal{J}_0(K_0)]}(\pm 1, 0)$, and the rate is independent of the hopping and given by $\gamma_{\text{lin}} = (\sqrt{g^2 + \omega^2} - g)[\mathcal{J}_2(K_0)/\mathcal{J}_0(K_0)](g/\omega)$.

Because $\mathbf{q}_{\text{mum}}^{\text{lin}}$ is located at the side edge of the 2D Brillouin zone, the saturation frequency for linear drive $\omega_c^{\text{lin}}$ equals half the effective 2D Bogoliubov bandwidth.

*Diagonal drive.*—Similarly, we find

$$\varepsilon_{\mathbf{q}}^{\text{diag}}(t) = 4J\left[\sin\frac{q_x}{2}\sin\left(\frac{q_x}{2} - K_0\sin\omega t\right) \right.$$
$$\left. + \sin^2\frac{q_y}{2}\sin\left(\frac{q_y}{2} - K_0\sin\omega t\right)\right],$$

$$h_{\mathbf{q}}^{\text{diag}}(t) = 4J\cos(K_0\sin\omega t)\left[\sin^2\frac{q_x}{2} + \sin^2\frac{q_y}{2}\right] - \varepsilon_{\text{eff}}(\mathbf{q})$$

$$= 8J\left(\sin^2\frac{q_x}{2} + \sin^2\frac{q_y}{2}\right)\sum_{l=1}^{\infty}\mathcal{J}_{2l}(K_0)\cos(2l\omega t).$$

(E6)

It follows that

$$s_{\text{diag}}(\mathbf{q}) = 4J\left(\sin^2\frac{q_x}{2} + \sin^2\frac{q_y}{2}\right)\mathcal{J}_2(K_0)\frac{g}{\omega}.$$

As expected, the expression is symmetric with respect to exchanging $q_x$ and $q_y$. The critical saturation frequency $\omega_c^{\text{diag}}$ is achieved when the maximally unstable mode reaches the Brillouin zone boundary $(q_x^{\text{mum}}, q_y^{\text{mum}}) = (\pi, \pi)$. For $\omega \geq \omega_c^{\text{diag}}$, the maximally unstable mode $\mathbf{q}_{\text{mum}}^{\text{diag}} = (\pi, \pi)$ and $\gamma_{\text{diag}} = 8J\mathcal{J}_2(K_0)g/\omega$. Notice that $\gamma_{\text{diag}} = 2\gamma_{\text{lin}}$ in this regime, due to the presence of the second shaken direction which enhances the instability rate. In the other regime, $\omega \leq \omega_c^{\text{diag}}$, we find $\mathbf{q}_{\text{mum}}^{\text{diag}} = 2\arcsin\sqrt{[\sqrt{g^2+\omega^2} - g/8J\mathcal{J}_0(K_0)]} \times (\pm 1, \pm 1)$ and $\gamma_{\text{diag}} = (\sqrt{g^2+\omega^2} - g)[\mathcal{J}_2(K_0)/\mathcal{J}_0(K_0)](g/\omega)$. Here, $\gamma_{\text{diag}} = \gamma_{\text{lin}}$.

Because $\mathbf{q}_{\text{mum}}^{\text{diag}}$ is located at the corner of the 2D Brillouin zone, the saturation frequency for diagonal drive $\omega_c^{\text{diag}}$ equals the full effective 2D Bogoliubov bandwidth.

*Circular drive.*—We find

$$\varepsilon_{\mathbf{q}}^{\text{circ}}(t) = 4J\left[\sin\frac{q_x}{2}\sin\left(\frac{q_x}{2} - K_0\sin\omega t\right) \right.$$
$$\left. + \sin^2\frac{q_y}{2}\sin\left(\frac{q_y}{2} + K_0\sin\omega t\right)\right],$$

$$h_{\mathbf{q}}^{\text{circ}}(t) = 4J\left[\sin^2\frac{q_x}{2}\cos(K_0\sin\omega t) \right.$$
$$\left. + \sin^2\frac{q_y}{2}\cos(K_0\cos\omega t)\right] - \varepsilon_{\text{eff}}(\mathbf{q})$$

$$= 8J\left(\sin^2\frac{q_x}{2} - \sin^2\frac{q_y}{2}\right)\sum_{l=1}^{\infty}\mathcal{J}_{2l}(K_0)\cos(2l\omega t).$$

(E7)

It follows that

$$s_{\text{circ}}(\mathbf{q}) = 4J\left|\sin^2\frac{q_x}{2} - \sin^2\frac{q_y}{2}\right|\mathcal{J}_2(K_0)\frac{g}{\omega}.$$

Note the change in the signature compared to the diagonal drive. Therefore, unlike the diagonal drive, and rather similar to the linear drive, the critical saturation frequency $\omega_c^{\text{circ}}$ is achieved when the maximally unstable mode reaches the side edge of the Brillouin zone $(q_x^{\text{mum}}, q_y^{\text{mum}}) = \{(\pi, 0), (0, \pi)\}$. For $\omega \geq \omega_c^{\text{circ}}$, the maximally unstable modes are $\mathbf{q}_{\text{mum}}^{\text{circ}} = \{(\pi, 0), (0, \pi)\}$ and $\gamma_{\text{circ}} = 4J\mathcal{J}_2(K_0)g/\omega$. Notice that $\gamma_{\text{diag}} = 2\gamma_{\text{circ}}$ and $\gamma_{\text{circ}} = \gamma_{\text{lin}}$ in this regime, due to the circular form of the drive. In the other regime, $\omega \leq \omega_c^{\text{circ}}$, we find $\mathbf{q}_{\text{mum}}^{\text{circ}} = 2\arcsin\sqrt{[\sqrt{g^2+\omega^2} - g/4J\mathcal{J}_0(K_0)]} \times \{(\pm 1, 0), (0, \pm 1)\}$ and $\gamma_{\text{circ}} = (\sqrt{g^2 + \omega^2} - g)[\mathcal{J}_2(K_0)/\mathcal{J}_0(K_0)](g/\omega)$. Here, $\gamma_{\text{diag}} = \gamma_{\text{circ}} = \gamma_{\text{lin}}$.

Because $\mathbf{q}_{\text{mum}}^{\text{circ}}$ is again located at the side edge of the 2D Brillouin zone, the saturation frequency for circular drive $\omega_c^{\text{circ}}$ equals half the effective 2D Bogoliubov bandwidth (similar to a linear drive).

The above analysis is performed for a pure 2D lattice system. However, adding the transverse tube dimension is straightforward and does not change these conclusions [24]; cf. Fig. 10 for a numerical simulation in the presence of a tubelike transverse direction.

Finally, note that the instability rate $\gamma$ derived above corresponds to the growth rate of the bosonic operators $b_{\mathbf{q}}(t)$ and $b_{\mathbf{q}}^\dagger(t)$, which are not observable. Observables, consisting of bilinears of the bosonic operators, will thus display a parametric instability rate

$$\Gamma_{\text{mum}} = 2\gamma \tag{E8}$$

for each maximally unstable mode.

Furthermore, an atom number variation per unit time (an experimental rate) depends on how many such modes are present simultaneously. For the linear drive, there exists a single maximally unstable mode $(\pi, 0)$, while there exist two equally unstable modes for the diagonal drive $[\{(\pi, \pi), (-\pi, \pi)\}]$ and for the circular drive $[\{(\pi, 0), (0, \pi)\}]$. Therefore, we expect an additional factor of 2 in $\Gamma_{\text{mum}}$ for the two 2D drives. In total,

$$\Gamma_{\text{mum}}^{\text{lin}} = 2\gamma_{\text{lin}}, \tag{E9}$$

$$\Gamma_{\text{mum}}^{\text{diag}} = 4\gamma_{\text{diag}}, \tag{E10}$$

$$\Gamma_{\text{mum}}^{\text{circ}} = 4\gamma_{\text{circ}}. \tag{E11}$$

As a final remark, we note that, while a choice of small $K_0$ explores the 2D quasiparabolic dispersion found at small BEC momenta, the above derivation shows that the problem still cannot be treated as rotationally symmetric: The resonant modes are determined by the drive frequency through the resonance condition. Therefore, even for small drive amplitude $K_0$, resonant Bogoliubov modes can be in the middle of the dispersion, where rotational symmetry is lost. Hence, the distinctions between 1D and 2D drives are expected, and observed, to hold even in the small $K_0$ limit.